# Interfacial thermal resistance between porous layers: impact on thermal conductivity of a multilayered porous structure

Pavlo Lishchuk[1c], Anastasiya Dekret[1], Anton Pastushenko[2,3], Andrey Kuzmich[1], Roman Burbelo[1], Ali Belarouci[2], Vladimir Lysenko[2] and Mykola Isaiev[1]

*(1) Faculty of Physics, Taras Shevchenko National University of Kyiv, 64/13, Volodymyrska Str., Kyiv 01601, Ukraine*

*(2) Université de Lyon; Institut des Nanotechnologies de Lyon, UMR-5270, site INSA de Lyon Villeurbanne F-69621, France*

*(3) Apollon Solar, 66 cours Charlemagne, Lyon 69002, France*

[C]*Electronic mail:* pavel.lishchuk@gmail.com

mykola.isaiev@knu.ua

## Abstract

Features of thermal transport in multilayered porous silicon nanostructures are considered. Such nanostructures were fabricated by electrochemical etching of monocrystalline Si substrates by applying periodically changed current density. Hereby, the multilayered structures with specific phononic properties were formed. Photoacoustic (PA) technique in gas-microphone configuration was applied for thermal conductivity evaluation. Experimental amplitude-frequency dependencies were adjusted by temperature distribution simulation with thermal conductivity of the multilayered porous structure as a fitting parameter. The experimentally determined values of thermal conductivity were found to be significantly lower than theoretically calculated ones. Such difference was associated with the presence of thermal resistance at the interfaces between porous layers with different porosities arising because of elastic parameters mismatch (acoustical mismatch). Accordingly, the magnitude of this interfacial thermal resistance was experimentally evaluated for the first time. Furthermore, crucial impact of the resistance on thermal transport perturbation in a multilayered porous silicon structure was revealed.

## Introduction

For decades, the drive for faster, cheaper computing and its associated device miniaturization have served to push scientists and engineers to develop materials, tools, processes, and design methodologies. State of the art electronic devices already operates with critical dimensions in the tens of nanometers. Development of the next generations of integrated circuits (ICs), three-dimensional (3D) integration and ultra-fast high-power density transistors has led to a steep increase in microprocessor chip heat flux and growing concern over the emergence of on-chip hot spots. Understanding thermal transport at the nanoscale is therefore crucial for a fundamental description of energy flow in nanomaterials, as well as a critical issue toward achieving optimal performance and reliability of modern electronic, optoelectronic, photonic devices and systems. Thermal transport at the nanoscale is fundamentally different from that at the macroscale and is determined by the distribution of carrier mean free paths and energy dispersion in a material, the dimension of the structure and the distance over which heat is propagated. The opportunity to shape new nanostructures that efficiently scatter phonons, reducing the thermal conductivity, without altering the electrical properties of the material enables the potential implementation of proficient thermoelectric devices to work as coolers or power generators from waste heat [1–4]. Recently, there has been much interest in the thermal conductivity of semiconductor superlattices (SLs) due to their promising applications in a variety of devices.

Since important surface-to-volume fraction in nanostructured materials, heat conduction across solid-solid interface predominates thermal transport there. Particularly, in SLs, multiple interfaces between different materials play a critical role in the thermal conductivity reduction [5,6]. Cutting-edge experimental techniques have enabled the measurements of the in-plane [7] and cross-plane [8] thermal conductivity in SLs. Experiments revealed that the thermal conductivity of SLs is strongly anisotropic and significantly lower than that estimated from the bulk value of the constituent materials and even smaller than the thermal conductivity of the equivalent composition alloys [9–12]. Theoretical investigation have established that the diffuse interface scattering induces the decrease of the in-plane (and, in part, the cross-plane) thermal conductivity, while the thermal boundary resistance (TBR) between adjacent layers is a key factor in the cross-plane thermal-conductivity reduction [13,14].

It is well-known, that stable mesoporous silicon thin films have thermal conductivity values two orders of magnitude lower than the thermal conductivity of bulk crystalline silicon [15,16]. Three orders of magnitude reduction has recently been achieved by applying additional treatment [17]. Thus, porous silicon is a promising candidate as a heat insulating material for micro- and nanoelectronic devices, especially taking into account its full compatibility with CMOS processing. Accordingly, systematical study of the heat transport in various porous silicon systems can give physical insight for further improvement of thermal engineering in silicon based elements.

In this article, we propose to study thermal properties of porous silicon based multilayered nanostructures. The aim of our work is to investigate thermal resistances of the interfaces between layers of different porosities. Photoacoustic (PA) technique in a classical configuration was used to probe the effective room temperature thermal conductivity across the multilayered porous nanostructures. Experimental analysis of thermal transport in the considered structure allows us to estimate interfacial thermal resistance values for different ratio of porosity between successive layers.

## Methods

*Experimental details*

Porous silicon multilayer structures were prepared by electrochemical anodization of (100)-oriented $p^+$- type boron-doped (0.01 - 0.02 Ohm·cm) monocrystalline silicon (c-Si) wafer (initial wafer thickness $l_{Si}$ = 510μm), which were previously immerged in a hydrofluoric acid (HF) solution to remove any native oxide. The electrochemical etching was carried out in an electrolyte based on 49 % HF and pure ethanol (at a volumetric ratio of 1:2). Anodization current density sequences (programmed by a Keithley DC power supply) corresponding to values in the range 7.5-80 mA/cm$^2$ was applied (see Table 1 in Appendix). The etched area was 1.0 cm$^2$, the total number of porous bi-layers was 30. Porosity values controlled by the etching conditions have been extracted from preliminary established calibration curves. Thicknesses of the etched porous layers have been measured from the scanning electron microscopy (SEM) images obtained by a MIRA 3 Tescan microscope. The cross-sectional SEM images of samples 1 and 2 are shown, for example, in Fig. 1.

Investigation of thermal transport properties was performed by applying the PA technique in classical configuration [17,18], at room temperature. All samples placed in the PA cell were irradiated by a laser diode module with an output optical power of about 60 mW (see Fig.2) and a $\lambda$ = 405 nm spectral wavelength. The light with an electrically modulated intensity was uniformly distributed over the whole sample surface by an optical system. The PA signal was detected by a microphone coupled to a lock-in nanovoltmeter in the 15-1000 Hz frequency range and compared to the reference signal delivered by the current generator.

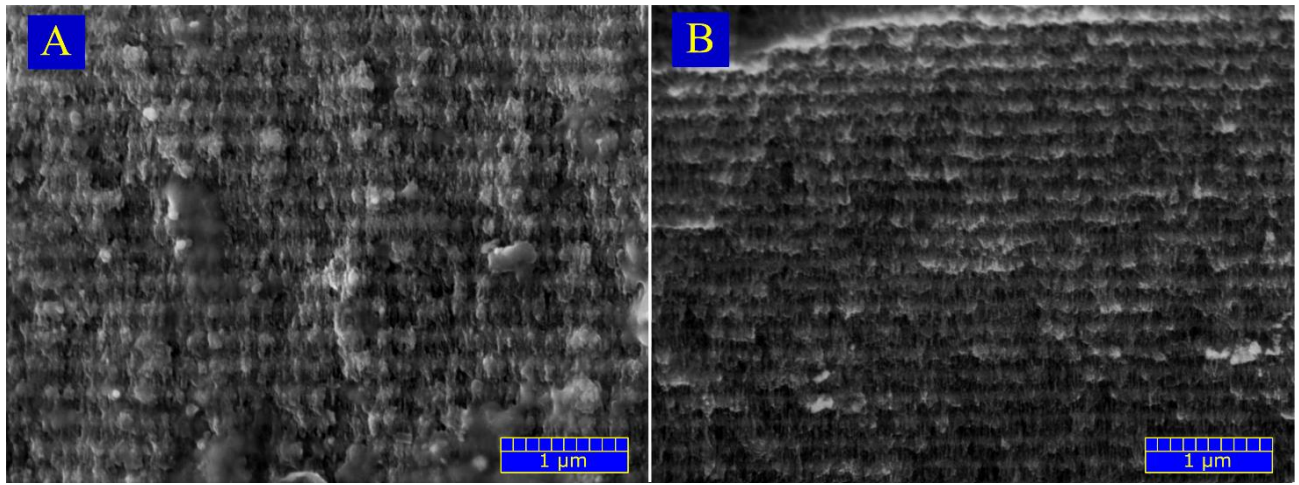

Figure 1. Cross-sectional SEM images of porous silicon multilayers of samples 1 (A) and 2 (B).

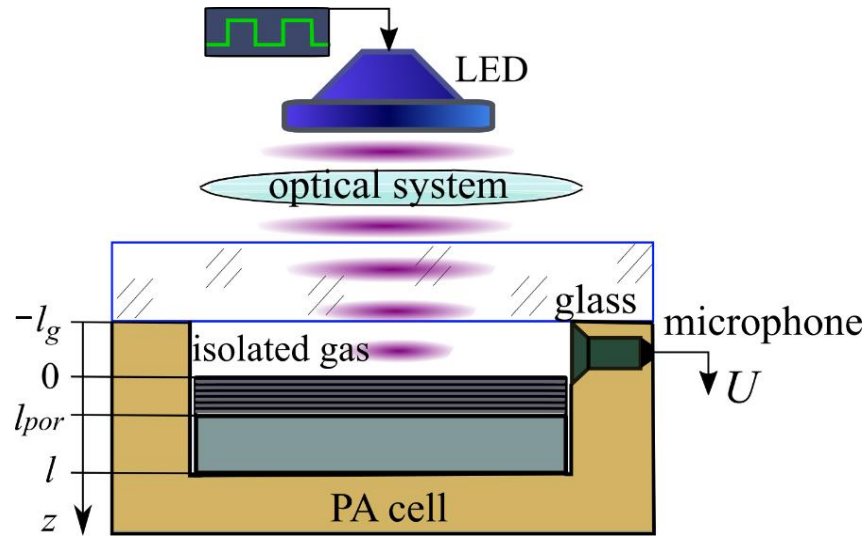

Figure 2. Experimental PA set-up in classic configuration.

As an example, experimental amplitude-frequency characteristics (AFC) of porous samples 5, 7, 14, 17, and the reference monocrystalline Si are presented in Fig. 3.

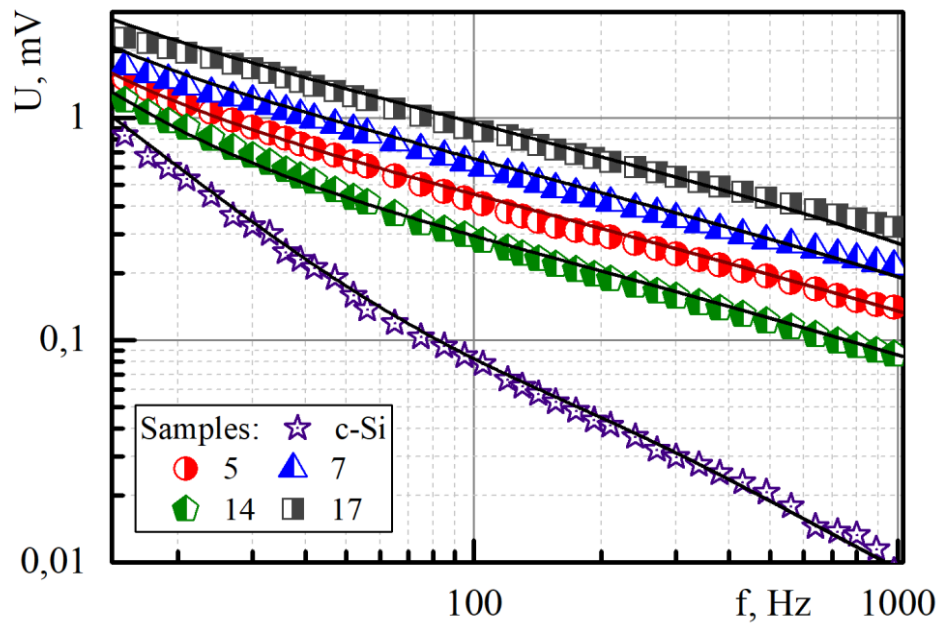

Figure 3. Experimental amplitude-frequency dependencies of the PA signal for samples 5, 7, 14, 17, and the reference monocrystalline Si. The solid lines

correspond to the calculated AFC obtained from the temperature distribution simulations described below.

*Simulation details*

Since the exciting UV light is absorbed within the first porous monolayer, its absorption coefficient (α) can be calculated from the Bruggeman's model [19,20] based on an effective medium approximation. Therefore, the permittivity of the porous monolayer $\varepsilon_{por}$ can be calculated as a function of porosity ($\xi$) with the following equation:

$$\xi \frac{1 - \varepsilon_{por}}{1 + 2\varepsilon_{por}} + (1 - \xi) \frac{\varepsilon_{Si} - \varepsilon_{por}}{\varepsilon_{Si} + 2\varepsilon_{por}} = 0 \qquad (1)$$

where $\varepsilon_{por}$ and $\varepsilon_{Si}$ are the complex permittivity of porous and bulk silicon, respectively. The well-known complex permittivity value of the bulk monocrystalline silicon can be found in the literature [21,22]. Absorption coefficient $\alpha_{por}$ of the porous layer can be deduced as follow (see Table 2 in Appendix):

$$\alpha = 4\pi k/\lambda \qquad (2)$$

where $k = Im(\sqrt{\varepsilon_{por}})$ is the imaginary part of the complex refractive index.

One-dimension model of photoacoustic transformation in a bi-layer structure has been used to determine the effective thermal conductivity of the porous samples [18]. In this model, the spatial distribution of the variable component of temperature $\theta(z)$ in the structure is induced by its periodic heating generated by light irradiation and is described by heat conduction equations with the corresponding boundary conditions:

$$\begin{cases} \frac{d^2\theta}{dz^2} - \frac{i\omega c_{por}\rho_{por}}{\chi_{por}}\theta = -\frac{I\alpha}{\chi_{por}}\exp(-\alpha z) & 0 < z < l_{por} \\ \frac{d^2\theta}{dz^2} - \frac{i\omega c_{Si}\rho_{Si}}{\chi_{Si}}\theta = 0 & l_{por} < z < l \end{cases},$$

$$\begin{cases} \left.\frac{\partial\theta}{\partial z}\right|_{z=0} = 0 \\ \theta_{z=l_{por}+0} = \theta_{z=l_{por}-0} \\ \chi_{por}\left.\frac{\partial\theta}{\partial z}\right|_{z=l_{por}-0} = \chi_{Si}\left.\frac{\partial\theta}{\partial z}\right|_{z=l_{por}+0} \\ \left.\frac{\partial\theta}{\partial z}\right|_{z=l} = 0 \end{cases} \quad (3)$$

where $\omega = 2\pi f$, $f$ is the modulation frequency of the light source, $\alpha$ is the optical absorption coefficient of the top porous layer, $I$ is the absorbed light intensity, $\chi_{por}$, $c_{por}$, $\rho_{por}$, are the thermal conductivity, heat capacity, and volume density of the array of porous silicon layers, respectively; $\chi_{Si}$, $c_{Si}$, $\rho_{Si}$ are the thermal conductivity, heat capacity, and mass density of the c-Si substrate, respectively; $l_{por}$ is the thickness of the porous layer, $l$ is the thickness of the entire structure.

The solution of the equations (3) could be written in the following form:

$$\theta = \begin{cases} Ae^{-\sigma_{por}z} + Be^{+\sigma_{por}z} - \frac{I\alpha}{\chi_{por}(\alpha^2-\sigma_{por}^2)}exp(-\alpha z) & 0 < z < l_{por} \\ De^{-\sigma_{Si}z} + Ee^{+\sigma_{Si}z} & l_{por} < z < l \end{cases} \quad (4)$$

where the complex constants $A$, $B$, $D$, $E$ can be numerically obtained from the boundary conditions, $\sigma_{por,Si} = \sqrt{i\omega c_{por,Si}\rho_{por,Si}/\chi_{por,Si}}$, respectively.

Pressure fluctuations ($P$) recorded by the microphone inside the PA cell can be evaluated with the Rosencsweig-Gersho model [23] and presented in the following form:

$$P(\omega) \sim \int_0^{-\infty} \theta(0) \cdot \exp(\mu_g z)\, dz \qquad (5)$$

where $\mu_g = \sqrt{i\omega c_g \rho_g / \chi_g}$; $\chi_g$, $c_g$, $\rho_g$ are the thermal conductivity, heat capacity and density of the isolated gas (air) respectively.

The thermal conductivity of porous layers was used as a fitting parameter to achieve the best correlation between the experimental and calculated data. The curves simulated with the best fitting parameters are presented in Fig. 3 as solid lines.

## Results and discussions

The estimated values of thermal conductivity ($\chi_{ex}$) for samples 1-9 are presented by filled circles in Fig. 4 as a function of the multilayer' period. These samples differ only by thicknesses of layer 1 ($l_1$) and 2 ($l_2$) while their porosities $\xi_1$ and $\xi_2$ were kept the same. As can be seen from the figure, the overall thermal conductivity of the multilayer nanostructures tends to a slight increase along with the increasing period ($l_1 + l_2$).

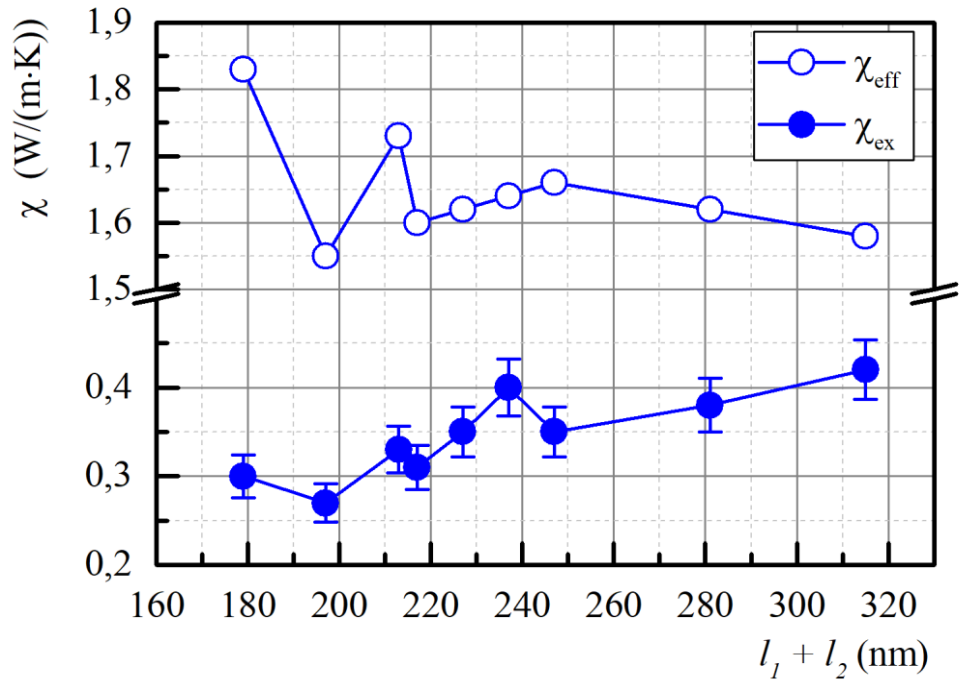

Figure 4. Experimental (filled circles) and effective (empty circles) thermal conductivities evaluated for sample 1 to 9.

The assessed thermal conductivities ($\chi_{ex}$) were compared to the effective ones ($\chi_{eff}$) calculated for the bi-layer (layers 1 and 2 with the thermal resistances $R_1 = l_1/\chi_1$ and $R_2 = l_2/\chi_2$, respectively) in the framework of an electro-thermal analogy model (as sketched in Fig. 5):

$$\frac{l_1 + l_2}{\chi_{eff}} = \frac{l_1}{\chi_1} + \frac{l_2}{\chi_2} \qquad (6)$$

where $\chi_1$ and $\chi_2$ are the thermal conductivity of layer 1 and 2, respectively. $\chi_1$ and $\chi_2$ have been measured on single porous silicon layer under the same etching conditions of layer 1 and 2, respectively [24,25] (see Table 2 in Appendix). Effective thermal conductivities calculated for samples 1-9 are presented in Fig. 4 with empty circles.

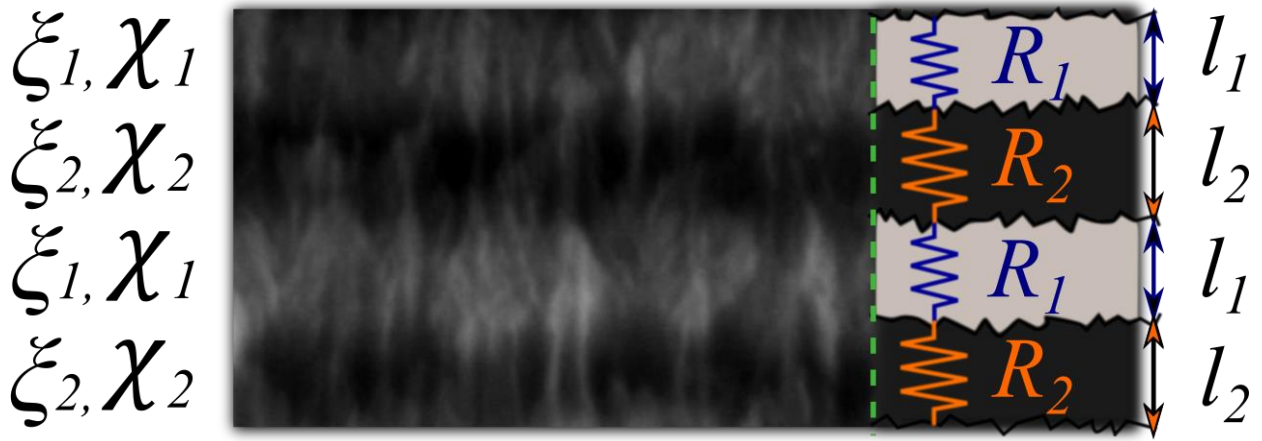

Figure 5. Sketch representation of the electro-thermal analogy model for the evaluation of the multilayers' effective thermal conductivity according to eq. (4).

As one can see in Fig. 4., the calculated effective thermal conductivities are found to be systematically higher in comparison with the experimental values. Moreover, the difference between the experimental and calculated values decreases with the increasing structural period $(l_1 + l_2)$. Such typical behavior can arise from an unaccounted boundary thermal resistance at interfaces between elementary porous layers of the multilayered nanostructure. Modification of Eq. (4) considering this effect can be re-written as follows:

$$\frac{l_1+l_2}{\chi'_{eff}} = \frac{l_1}{\chi_1} + \frac{l_2}{\chi_2} + 2R \qquad (7)$$

where $R$ is the boundary thermal resistance. This value can be easily extracted from equation (5) considering that: $\chi'_{eff} = \chi_{ex}$. The thermal resistance value for samples 1-9 (see the corresponding data in Table 1) was found to be: $(2.7 \pm 0.3)\,10^{-7}$ (m² K)/W. This value is in good agreement with the Kapitza thermal resistance at solid-solid interfaces [26,27].

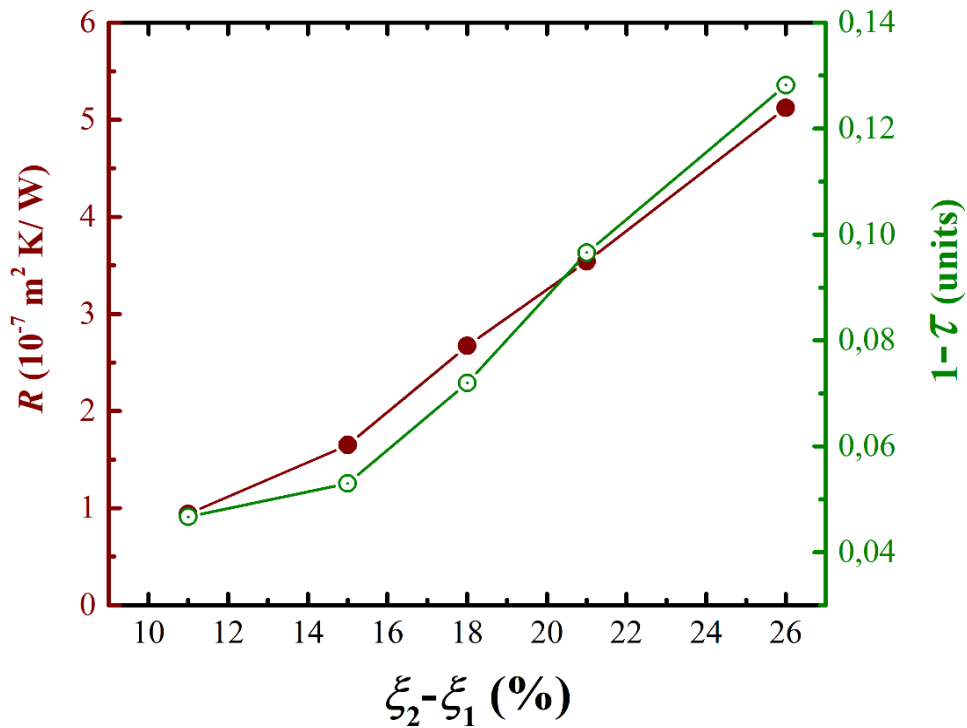

Figure 6. Interfacial thermal resistance and interface reflectivity dependencies as a function of porosity difference between adjacent layers.

The same approach was used for the determination of thermal resistances for all other studied samples. The resulting dependence of the boundary thermal resistances on porosity difference ($\xi_2 - \xi_1$) between layers 1 and 2 are presented by filled circles in Fig. 6. As one can see, this dependence is clearly characterized by a quasi-linear function. Such a trend is related to a difference between sound velocities in layers of different porosities. The latter statement correlates with an acoustical mismatch model for a multilayered structure. To check this fact, we have considered the phonon transmission ($\tau$) of an interface in the framework of this model [28]:

$$\tau = \frac{4 z_1 z_2}{(z_1 + z_2)^2},\quad (8)$$

where $z_i = \rho_i v_i$ is the acoustical impedance of the i-th layer, $\rho_i$ is the mass density of the i-th layer, and $v_i$ is the phonon velocity in the i-th layer.

For evaluation of the interface transmission, the phonon velocities in porous silicon prepared under the same condition have been used [29] (see Table 2 in Appendix). A phonon propagation at normal incidence on interfaces between elementary porous layers can be assumed. This assumption is quite acceptable considering nanoscale sizes (10-20 nm) of Si crystallites and phonon mean free path in bulk silicon (43 nm at room temperature [14]). Evolution of the interface reflectivity ($1 - \tau$) as a function of the porosity difference ($\xi_2 - \xi_1$) is also presented in Figure 6 by empty circles. As one can see, a perfect correlation of this dependence with the evolution of the boundary thermal resistance can be stated.

## Conclusion

The PA technique in classic gas-microphone configuration was used to study the heat transport in multilayered porous silicon nanostructures. Experimental amplitude-frequency dependencies of the PA signal were also theoretically simulated to evaluate thermal conductivities of the nanostructures. The experimental values are significantly lower than those estimated in the framework of an electro-thermal analogy model. The difference between the experimental and theoretical values of thermal conductivity have been associated with the presence of boundary thermal resistance at the interfaces between porous monolayers. A typical increasing linear dependence of the thermal boundary resistance with porosity difference was found. The latter corresponds well to the acoustical mismatch model for the Kapitza thermal resistance.

# Appendix

Table 1

Fabrication and morphology characteristics of the investigated porous silicon multilayered samples

| Sample | $t_1$, s | $t_2$, s | $j_1$, mA/cm$^2$ | $j_2$, mA/cm$^2$ | $\xi_1$, % | $\xi_2$, % | $l_1$, nm | $l_2$, nm |
|---|---|---|---|---|---|---|---|---|
| 1  | 6  | 4 | 10   | 50 | 46 | 64 | 61  | 136 |
| 2  | 8  | 4 | 10   | 50 | 46 | 64 | 81  | 136 |
| 3  | 9  | 4 | 10   | 50 | 46 | 64 | 91  | 136 |
| 4  | 10 | 4 | 10   | 50 | 46 | 64 | 101 | 136 |
| 5  | 11 | 2 | 10   | 50 | 46 | 64 | 111 | 68  |
| 6  | 11 | 3 | 10   | 50 | 46 | 64 | 111 | 102 |
| 7  | 11 | 4 | 10   | 50 | 46 | 64 | 111 | 136 |
| 8  | 11 | 5 | 10   | 50 | 46 | 64 | 111 | 170 |
| 9  | 11 | 6 | 10   | 50 | 46 | 64 | 111 | 204 |
| 10 | 10 | 4 | 5    | 50 | 42 | 64 | 62  | 136 |
| 11 | 10 | 4 | 7.5  | 50 | 44 | 64 | 82  | 136 |
| 12 | 10 | 4 | 15   | 50 | 48 | 64 | 138 | 136 |
| 13 | 10 | 4 | 17.5 | 50 | 50 | 64 | 155 | 136 |
| 14 | 10 | 4 | 10   | 30 | 46 | 57 | 101 | 93  |
| 15 | 10 | 4 | 10   | 40 | 46 | 61 | 101 | 115 |
| 16 | 10 | 4 | 10   | 60 | 46 | 67 | 101 | 152 |
| 17 | 10 | 4 | 10   | 70 | 46 | 71 | 101 | 169 |
| 18 | 10 | 4 | 10   | 80 | 46 | 72 | 101 | 192 |

where: $t_1$, $t_2$ – etching times, $j_1$, $j_2$ – current densities, $\xi_1$, $\xi_2$ – porosities, $l_1$, $l_2$ – thicknesses of the layers 1 and 2, respectively.

Table 2

Fitting parameters of the investigated porous silicon multilayered samples

| Sample | $\chi_1$, W/(m K) | $\chi_2$, W/(m K) | $\chi_{eff}$, W/(m K) | $\alpha$, $10^4$ cm$^{-1}$ | $\chi_{ex}$, W/(m K) | $R$, $10^{-7}$ m$^2$ K/W | $c_1$, $10^3$ m/c | $c_2$, $10^3$ m/c |
|---|---|---|---|---|---|---|---|---|
| 1 | 2.36 | 1.34 | 1.55 | 7.95 | 0.27 | 3.01 | 5.2 | 4.5 |
| 2 | 2.36 | 1.34 | 1.6 | 7.95 | 0.31 | 2.82 | 5.2 | 4.5 |
| 3 | 2.36 | 1.34 | 1.62 | 7.95 | 0.35 | 2.54 | 5.2 | 4.5 |
| 4 | 2.36 | 1.34 | 1.64 | 7.95 | 0.4 | 2.24 | 5.2 | 4.5 |
| 5 | 2.36 | 1.34 | 1.83 | 7.95 | 0.3 | 2.49 | 5.2 | 4.5 |
| 6 | 2.36 | 1.34 | 1.73 | 7.95 | 0.33 | 2.61 | 5.2 | 4.5 |
| 7 | 2.36 | 1.34 | 1.66 | 7.95 | 0.35 | 2.79 | 5.2 | 4.5 |
| 8 | 2.36 | 1.34 | 1.62 | 7.95 | 0.38 | 2.83 | 5.2 | 4.5 |
| 9 | 2.36 | 1.34 | 1.58 | 7.95 | 0.42 | 2.75 | 5.2 | 4.5 |
| 10 | 2.75 | 1.34 | 1.6 | 8.53 | 0.3 | 2.68 | 5.42 | 4.5 |
| 11 | 2.52 | 1.34 | 1.63 | 8.27 | 0.25 | 3.69 | 5.48 | 4.5 |
| 12 | 2.08 | 1.34 | 1.63 | 7.63 | 0.5 | 1.90 | 5.06 | 4.5 |
| 13 | 1.94 | 1.34 | 1.6 | 7.33 | 0.55 | 1.74 | 4.82 | 4.5 |
| 14 | 2.36 | 1.52 | 1.87 | 7.95 | 0.66 | 0.95 | 5.2 | 4.68 |
| 15 | 2.36 | 1.4 | 1.73 | 7.95 | 0.55 | 1.34 | 5.2 | 4.3 |
| 16 | 2.36 | 1.28 | 1.57 | 7.95 | 0.25 | 4.25 | 5.2 | 4.55 |
| 17 | 2.36 | 1.23 | 1.5 | 7.95 | 0.28 | 3.92 | 5.2 | 4.68 |
| 18 | 2.36 | 1.22 | 1.46 | 7.95 | 0.2 | 6.32 | 5.2 | 4.64 |

where: $\chi_1, \chi_2$ – thermal conductivity of the porous monolayers 1 and 2, respectively (from Ref. [24, 25]), $\chi_{eff}$ – the effective thermal conductivity, $\alpha$ – optical absorption coefficient of the top photoexcited porous layer, $\chi_{ex}$ – experimental thermal conductivity value, R – boundary thermal resistance, $c_1$, $c_2$ – values of phonon velocities (from Ref. [29]).